# Benchmark Results and Theoretical Treatments for Valence-to-Core X-ray Emission Spectroscopy in Transition Metal Compounds

D.R. Mortensen[1,2], G.T. Seidler[1, (*)], Joshua J. Kas[1], Niranjin Govind[3], Craig P. Schwartz[4,5], Sri Pemmaraju[4], David G. Prendergast[4]

[1] Physics Department, University of Washington, Seattle, WA 98195, USA
[2] easyXAFS LLC, Seattle, WA 98122, USA
[3] Environmental Molecular Sciences Laboratory, Pacific Northwest National Laboratory, Richland, WA 99352, USA
[4] The Molecular Foundry, Lawrence Berkeley National Laboratory, Berkeley, CA 94720, USA
[5] Stanford Synchrotron Radiation Lightsource, SLAC National Accelerator Laboratory, Menlo Park, CA 94025, USA

We report measurement of the valence-to-core (VTC) region of the K-shell x-ray emission spectra from several Zn and Fe inorganic compounds, and their critical comparison with several existing theoretical treatments. We find generally good agreement between the respective theories and experiment, and in particular find an important admixture of dipole and quadrupole character for Zn materials that is much weaker in Fe-based systems. These results on materials whose simple crystal structures should not, a priori, pose deep challenges to theory, will prove useful in guiding the further development of DFT and time-dependent DFT methods for VTC-XES predictions and their comparison to experiment.

## I. Introduction

In the current landscape, the field of x-ray absorption spectroscopy (XAS) occupies a position of broad scientific scope and technological importance. This footing, however, was not easily achieved. While the roots of XAS extend back to the first observations by de Broglie in 1913,[1] the first 60 years of its life was spent as a topic of fundamental research with limited opportunity for application. It was not until the 1970s, with the establishment of several electron storage rings for dedicated synchrotron radiation experiments,[2-4] that XAS became a methodology with steadily growing reliability, availability and, especially, breadth of impact. Since that time a tremendous amount of work and resources have gone into building synchrotron lightsources, and more recently x-ray free electron lasers, around the world.[5-7]

The theoretical understanding of XAS has been similarly fraught. Settling the central conceptual issue of the *locality* of the interrogated density of states was a nearly 50 year battle.[8] The discovery of the 'EXAFS equation',[9] effectively casting the extended absorption oscillations as a single- or few-scattering process, was merely the first shot which launched several decades of work in finding optimal descriptions of the phase shifts due to atomic and inter-atomic potentials.[10-19] The establishment of reliable theoretical predictions and interpretations for oscillations in the main body of the near-edge fine structure required, first, a simplified computational framework for the influence of full-multiple scattering and, second, a careful treatment of core-hole effects.[8] While a number of these issues have been settled, others are still matters of contemporary research. Chief among these is the interpretation of pre-edge features, especially those coupled to charge-transfer effects or other dynamical rearrangement of charge density that go beyond 'typical' excitonic effects induced by the core-hole potential.[20-26]

The history of XAS hence represents a clear example of a reoccurring lesson in science: the growth of any analytical method requires parallel development in each of instrument technology, cross-technique validation, and theory. Indeed, due to the lack in each of the prior three criteria, few could have seriously imagined in the early 1970's that XAS would evolve to the point where it is now routinely used to solve forefront problems in metallorganic chemistry or that it would become a work-horse for industrial and fundamental research in catalysis[27-31] and electrical energy storage,[32-37] to name only a few prominent examples.

Following in the technical and, to a growing extent, historical footsteps of XAS, x-ray emission spectroscopy (XES) has over the past few years emerged as an important new spectroscopic tool, spreading from the realm of fundamental condensed matter science to, e.g., applications in catalysis,[28,38-43] electrochemistry,[44-46] biological sciences.[47-53] While the semi-core and deeper core transitions involved in XES are often reasonably well described by perturbed atomic multiplet approaches due to the extreme localization of the atomic-like initial and final states, the situation is markedly less clear for those transitions involving valence electron density of the host species and ligands.

As the name suggests, this valence-to-core (VTC) x-ray emission involves the filling of a deep core-hole via de-excitation of valence-level electronic states. The valence orbitals, with energies within a few eV to ~15 eV of the Fermi level, are the most sensitive to the chemical environment and therefore VTC-XES has much greater sensitivity to local coordination effects than do diagram lines involving only deeper core shells. While various other x-ray spectroscopy techniques exist (e.g., x-ray photoemission, x-ray absorption, x-ray Raman, *etc*.), there exist a number of fine issues of local and electronic structure that are best addressed through VTC-XES. A recent, well-known example is the identification the central atom in the nitrogenase iron-molybdenum cofactor for dinitrogen reduction in biological and industrial catalysis.[41] From the most general perspective, VTC-XES should be viewed as a natural complement to the pre-edge and very near-edge regions of XAS, in that VTC-XES is sensitive to the occupied, rather than unoccupied, states near the Fermi level. At the same time, VTC-XES comes with a certain advantage in that, due to the final-state rule, theoretical treatment of VTC-XES is simplified because of the absence of a core hole after emission.

Again, following the developmental history of XAFS, and all other modern spectroscopies, when the applications and demands of VTC-XES expand, so too must the supporting infrastructure in experimental apparatus and in methodology and *validation* of theory. While the early stages of growth in XES have benefitted from the pioneering work conducted at several synchrotron end-stations,[51,53-58] the relative scarcity of these dedicated beamlines is a serious hurdle to routine application. This has led to continuing effort by several groups to develop laboratory-based XES capabilities.[59-63] Here, using this equipment at the University of Washington,[64-66] we present a high-quality VTC dataset of several inorganic Zn and Fe compounds. These compounds provide an interesting range of local electronic and atomic structure while retaining sufficient structural simplicity such that theoretical treatment should not, *a priori*, be challenged by material complexity.

To date, the most successful models are those based on density functional theory (DFT). Different implementations, however, often differ in significant ways (treatment of electron exchange-correlation, basis sets, real vs. reciprocal space, inclusion of relativistic effects, etc.). We therefore present a critical assessment of several state-of-the-art DFT-based electronic structure codes in the context of this new experimental dataset. While the proper choice of theoretical method may vary from application to application, the present investigation will help identify strengths and limitations of the various approaches. We note that we are not the first to seek a better understanding of the validity of theory in order to expand the range of application of VTC-XES. In recent years, the DeBeer group and collaborators have embarked upon a course of study aimed at establishing the information content available in VTC-XES of complex molecular systems[38-43,52] with the ultimate goal of using time-dependent DFT to develop an understanding of chemical information in a molecular orbital framework.

This manuscript continues as follows. First, in section II, we present experimental details. This includes both sample preparation and details of the laboratory-based spectrometer used here. Second, in section III, we provide technical details for the implementation of the three different theoretical codes that are compared to experiment. Next, in section IV, we present results and discussion. This begins with necessary demonstration of baseline spectrometer performance metrics and the methods used for subtraction of fluorescence contributions not associated with the VTC transitions, subsequently continuing to a complete presentation of all experimental and theoretical results. We conclude in section V.

## II. Experimental

All samples for this study were prepared from high purity powders (99.9% or better) from Sigma Aldrich or Alfa Aesar, the exception being the Zn and Fe metal samples which were foils (99.9%) from ESPI Metals. Powder samples were pressed into few-mm thick pellets and encased in pouches made from 25-μm thick polyimide films.

Although VTC features were first observed in the laboratory as early as the mid-1930s,[67-69] it is only in recent years that laboratory-based equipment has been employed in chemical studies.[59] Here we employed a Rowland-circle spectrometer developed at the University of Washington.[64-66] This low-powered prototype instrument achieves synchrotron-quality energy resolution and also count rates comparable to what would be obtained for the same XES studies at monochromatized bending magnet beamlines at third-generation synchrotrons.

Briefly, sample fluorescence was stimulated sample via output from a commercial x-ray tube (MOXTEK, Inc.) operated at 40 kV accelerating potential with 200 µA electron beam current incident on an Au anode. Sample fluorescence was analyzed using 10-cm diameter, spherically bent Ge (555) and Ge (620) crystals for Zn and Fe respectively, each with a 1-m radius of curvature (XRS Tech, LLC). Analyzed x-rays are detected with a silicon drift detector (Amptek, Inc.) and a region of interest of a few hundred eV wide was set to strongly reject background signal. Data was recorded in 0.25-eV steps with 30-50 s of integration per point in the $K\beta_{1,3}$ region and 100-160 s/point for the VTC. For the Fe (Zn) results, each Fe (Zn) XES spectrum is on the same energy scale to high precision[65] and a single overall shift of energy scale is used to calibrate with respect to published values for the $K\beta_{1,3}$ peak location of Fe (Zn).

We note that the spectral resolution is poorer for the Fe compounds than it is for Zn (where it is close to core-hole limited). We believe this result stems from defects in the Ge (620) optic leading to increased bandwidth. Nonetheless, the performance is sufficient to cleanly resolve key features in the VTC spectra.

As DFT is ill-equipped to model the core-to-core $K\beta_{1,3}$ emission due to difficulties in correctly estimate 3p-3d splitting, the intensity contribution of the high-energy tails of these lines are typically subtracted from the valence region for comparison of theory to experiment. To this end, each full spectrum was fit to a series of pseudo-Voigt functions and a constant background using the *Blueprint XAS* package.[70,71] In addition to the main $K\beta_{1,3}$ and valence features, we include extra curves to model the multi-electron excitation peaks (KLβ) above the Fermi-level[72-74] and the radiative Auger satellites[73] in the intermediate area as such features are not accounted for in the base theories. For the Zn spectra an additional pseudo-Voigt function is included to model the elastically scattered Au Lα$_2$ line originating from the tube anode. The width and position of this curve was constrained to be consistent across all samples. To emphasize the valence region in fitting, it was assigned a weighting of 6:1 relative to the Kβ$_{1,3}$. Representative results of this procedure are shown in section IV.A, below.

## III. Theoretical Methods

We perform calculations using three state-of-the-art, *ab intio* electronic structure packages: Quantum ESPRESSO (QE),[75] FEFF,[76] and NWChem.[77] While each of these codes has a basis in DFT, they are built around distinct treatments leading to unique calculations. We briefly discuss the methodology for each implementation below.

First, calculations were performed within the generalized gradient approximation-DFT framework using ultra-soft pseudo potential with 125 Ry energy cutoff implemented in the QE package with adequate *k*-point sampling for convergence with the PBE correlation and exchange.[75,78] We calculate the off-resonant XES spectrum assuming the 'final-state rule' which assumes a filled core-hole and a screened valence-hole. The spectra calculated here consider only dipole contributions to the transitions and are thus due to *p*-type projection of the density of states (DOS). To simulate the natural core-hole lifetime broadening and experimental resolution, the calculated stick spectra were Lorentzian broadened by 6.0 eV and 2.5 eV for Fe and Zn respectively. Each spectrum was then shifted independently in energy to align with experiment. It should be noted that the calculated spectral widths tend to be unphysically compressed due to the well-known problem of DFT underestimating the band gap.[79]

Second, theoretical spectra were also simulated using a full multiple scattering method implemented within the FEFF 9.6 code.[76,80] The potentials were calculated self-consistently using a 5.0 Å cluster. The spectra were calculated using a full multiple scattering radius of 6.0 Å. The initial core-state energy levels were calculated using the final self-consistent potential. This modification is intended to provide more accurate relative chemical shifts. Here, both electric dipole and quadrupole transitions were included. Following the standard practice for XES, these calculations were performed with no core hole. For comparison to experiment, these results were convolved with a Lorentzian (4.5 eV for Fe and 0.5 eV for Zn) and shifted independently in energy to match experiment. FEFF also includes calculations for the main $K\beta_{1,3}$ lines, but these contributions have been removed for the sake of comparison.

Finally, the VTC-XES approach in NWChem is based on linear-response time-dependent density functional theory (LR-TDDFT), which has been used successfully to simulate the VTC-XES spectra of low- and high-spin model molecular complexes involving Cr, Mn and Fe transition metal centers in good comparison with experiment.[81] First a neutral ground state calculation is performed, a full core hole (FCH) ionized state is then obtained self-consistently where the 1s core orbital of the transition metal (TM) absorption center is swapped with a virtual orbital combined with the maximum overlap constraint to prevent core hole collapse. A LR-TDDFT calculation, within the Tamm-Dancoff approximation (TDA), is then performed with the FCH reference state to simulate the VTC emission process. This approach allows one to go beyond the single-particle picture as all orbital pairs with significant contributions to the emission process are included naturally. To describe excitations beyond the dipole approximation, higher-order contributions are included in the calculation of the oscillator strengths.

All systems (non-magnetic Zn compounds) were represented with finite clusters constructed from crystal structures obtained from experiment. To account for the surface states, the clusters were terminated using a set of suitably chosen pseudo-hydrogen saturators whose charges are calculated using the formal charges of the surface atoms.[82,83] The Los Alamos effective core potential (LANL2DZ)[84-87] and associated basis sets were used for all the atoms (Zn, Cl, S, O) except the Zn absorbing center in each system which was represented with the Sapporo-TZP-2012[88] all electron basis set. The PBE0 exchange-correlation functional[89] was used for all calculations. For comparison to experiment, each calculated spectrum was convolved with a 2.0 eV Lorentzian and energy shifted. We note that, in contrast to the QE and FEFF results, this shift was constant across all samples indicating an accurate accounting for chemical shifts. Unfortunately, the corresponding calculations for the Fe materials were not performed in this study due to the added complexity of dealing with magnetic effects in finite cluster calculations.

## IV. Results and Discussion

### IV.A. Instrument Baseline Performance

To begin, it is important to briefly consider instrument performance and its systematic limitations before proceeding to the results themselves. First, in Figure 1, we show a typical spectrum from Fe metal, with data collection extending well past the Fermi level. Note that the figure is presented on a semi-logarithmic scale. The key point is that the noise floor from stray scatter is far below the intensity of the VTC transition.

Second, in Figure 2, we show the instrumental insensitivity to sample preparation or positioning. This important characteristic, described in detail elsewhere,[65] is a consequence of moving the sample location slightly behind the Rowland circle and inserting an entrance slit onto the nominal 'source' location on the Rowland circle. The resulting spectra have less than 25 meV irreproducibility in overall energy scale even upon large sample movement or sample exchange.

Third, as mentioned in Section II, DFT methods do not calculate several real fluorescence 'backgrounds' that contribute in the same energy range, i.e., the high-energy tail of the $K\beta_{1,3}$ fluorescence, radiative Auger contributions, or fluorescence resulting from multi-electron excitations. Consequently, we follow prior practice and make use of physically-motivated fits to these backgrounds; a representative example is presented in Fig. 3. In the case of the Zn samples, the 'background' contributions from the $K\beta_{1,3}$ are nearly identical across all species. Minor 'background' variances occurred primarily in the intensity of a modest background peak due to Au elastic scatter line and in the shape of the multi-electron excitations appearing about the single-particle Fermi level. The latter is not unexpected, as the observed intensity of these lines are strongly influence by sample geometry and the structure of the absorption coefficient, as measured in XANES.[74,90-93]

The issue of sample re-absorption of fluorescence prior to escape is also important in the shape of the $K\beta_{2,5}$ emission peak as its high-energy side often straddles the rising K-edge. This creates an important systematic effect in thick samples; fluorescence above the absorption edge is preferentially quenched when escaping outward from the sample bulk. Due to the fine structure modulations in absorption, and in some cases strong pre-edge features, this effect often distorts spectral shape in significant ways. As an example, self-absorption causes the apparent asymmetry in the $K\beta_{2,5}$ peak of Fe shown in Figure 4.

In principle, sample self-absorption is correctable if the absorption coefficient, as measured in x-ray absorption spectroscopy, and sample thickness are known. An accurate correction, however, requires high precision in the relative energy scale between emission and absorption measurements. This is highly nontrivial as different instrumental setups are required for each type of measurement, and we do not attempt a correction for the data presented in this study, but a recent manuscript describes the methods needed for this correction in the context of multi-electron excitations in Ni metal.[93] Here, we consider only the performance of our calculations only below the nominal edge energies (7112 eV for Fe and 9659 eV for Zn).

**IV.B. Experimental Spectra and Comparison to Theory**

Our VTC XES spectra for Zn compound are shown in Figure 5. Note, in particular, the clear splitting between the K$\beta_2$ and K$\beta_5$ lines, a situation that is somewhat unique to Zn among the transition metals. In Fe (below), and indeed most 3d-transition metals, these two features are indistinct due to core hole broadening and are thus referred to together as K$\beta_{2,5}$. The origins of the weak K$\beta_5$ line, which was first investigated in the earlier twentieth century,[94] remains uncertain. While it is generally regarded as quadrupole-allowed transitions from state of metal 3d character,[95-98] it has recently been suggested that the major contribution could come instead from dipole allowed 4p-type states from neighboring atoms.[99]

To address this issue, we present in Figure 6 the electric dipole and quadrupole contributions to the VTC spectrum of ZnO, as determined by FEFF. As the K$\beta_5$ sits atop the tail of the K$\beta_2$ line, we isolate the K$\beta_5$ dipole contribution for an accurate comparison. These calculations suggest that the above interpretations of K$\beta_5$ origin are individually incomplete and that both terms significantly contribute to the overall intensity.

We return now to a comparison of the three calculations shown in Fig. 5. Outside of the missing quadrupole contribution in QE, we see similar predictions made between it and FEFF in terms of splitting between features, including a common underestimation of the splitting between K$\beta_2$ and K$\beta_5$. Such a compression is a well-known problem in DFT, arising from difficulties in correctly predicting the bandwidth.[79]

In contrast, LR-TDDFT based approach in NWChem generally shows an improved relative spacing of features compared to experiment. This response approach allows one to go beyond the single-particle picture as all orbital pairs with significant contributions (or multi-configurational character) to the emission process are naturally included. We note that unlike the calculations for QE (3.2 eV spread) and FEFF (6.3 eV spread), the NWChem predictions require a single, *consistent* energy shift across all samples to align with experiment. The ability to reliably predict relative energy shifts across sample chemistries is an important feature in VTC-XES analysis and hence this is a significant result in characterizing NWChem performance. One weakness in the NWChem results is the prediction of an apparent unphysical peak at ~9646 eV for pure Zn metal. We believe this feature may be an artifact of the finite cluster size used to represent a metallic system.

Next, we present the results for several Fe-rich samples in Figure 7, including comparison to QE and FEFF. Overall, we see good reproduction of the experiment by both theories. We note that the QE and FEFF calculations produce similar spectra with nearly identical splitting between the K$\beta_{2,5}$ and K$\beta''$ peaks, the latter being a cross-over feature originating from ligand orbital with metal-p character. The magnitude of this splitting, however, appears to be slightly underestimated especially in the case of $Fe_2O_3$ and $Fe_3O_4$. As with the Zn results, this is likely due to difficulties in correctly predicting the bandgap. In the case of FeS, QE misses the cross-over peak entirely, which we believe to be an issue of the DFT misidentifying the character of the state. In general, the intensity of this feature tends to be under-predicted with respect to FEFF.

Unlike with Zn, the good agreement between FEFF (which contains both dipole and quadrupole contributions) and QE (dipole contributions only) for the Fe materials suggests that quadrupole contributions are weaker for Fe. In Figure 8, we present a separation of each term in the FEFF output for $Fe_2O_3$, confirming this assertion. The intensity of the quadrupole features is similarly negligible across all Fe-rich samples.

Again, we stress that the required energy shifts to align calculation with experiment are inconsistent across samples, with a relative spread 3.5 eV and 10.6 eV for QE and FEFF respectively. As seen from this dataset the relative positioning of VTC features is not fixed, with real, physical shifts occurring due to changes in chemical state, particularly oxidation. This deficiency is therefore a topic that must be addressed in order to establish a robust, *ab initio* interpretation of experimental spectra.

The above results suggest several interesting results that can guide improved theoretical treatment and its comparison to experiment. First, the mixed dipole-quadrupole nature of the $K\beta_5$ feature is likely not specific to the present simple crystal structures and the question of the magnitude of possible quadrupole character of the $K\beta_5$ feature for other transition metal species should be considered, although it does appear to be weak for Fe in the present study. Second, and not unexpectedly, the compression of features due to underestimation of bandwidths will be a persistent issue in the treatment of VTC-XES with DFT, although some progress should be possible with, e.g., using a GW approximation for the quasiparticle energy shifts.[100,101] Finally, there do appear to be benefits to using time-dependent DFT-based approach. With the exception of pure Zn, where we believe cluster-size effects played a role, we see generally better agreement with the positions and amplitudes of the observed VTC-XES features in the Zn compounds.

Before concluding, it is useful to compare and contrast the present degree of agreement between theory and experiment with that observed in prior work conducted on molecular systems containing transition metals. In those studies, time-dependent DFT calculations were applied within the ORCA quantum chemistry package.[102] The full details of their methodology can be found elsewhere.[40,103] Overall, the strength and weaknesses of these calculations in reproducing experiment match well with our own results discussed above. In general, the ORCA-DFT results have been reported to successfully track relative intensities of VTC features.[39,40,49,52] While the absolute energy scale can deviate significantly, the relative energy scale, both between features within a single spectrum and when comparing chemical shifts across samples, tend to show excellent agreement much like the NWChem predictions.[40,52] In some cases, however, this code has been shown to predict features which are apparently absent in the experimental data, but that may be difficult to find in experiment due to limitations in removing the $K\beta_{1,3}$ background and due to the larger Poisson noise induced by this background.[52]

## V. Conclusions

In summary, we have presented a high-quality VTC-XES dataset of several inorganic Zn and Fe compounds. Using this dataset, we evaluated several state-of-the-art DFT-based electronic structure codes each built around distinct theoretical treatments. While each code showed generally good agreement with experiment, we find a number of important features (relative chemical shifts, importance of higher-order transitions, energy splitting, *etc.*) that distinguish their performance. These results should prove useful in guiding the further development of DFT and time-dependent DFT methods for VTC-XES and their comparison to experiment.

**Acknowledgements**

The work at the University of Washington was supported by the United States Department of Energy, Office of Science, under grant DE-SC0002194. The Stanford Synchrotron Radiation Lightsource and Molecular Foundry are National User Facilities operated by Stanford University and the University of California Berkeley for the US Department of Energy, grants DE-AC02-76SF00515 and DE-AC02-05CH11231 respectively. NG acknowledges support from the Chemical Sciences, Geosciences, and Biosciences Division, Office of Basic Energy Sciences, Office of Science, U.S. Department of Energy through Award KC030105066418. A portion of the research was performed using EMSL, a DOE Office of Science User Facility sponsored by the Office of Biological and Environmental Research and located at PNNL. PNNL is operated by Battelle Memorial Institute for the United States Department of Energy under DOE contract number DE-AC05-76RL1830

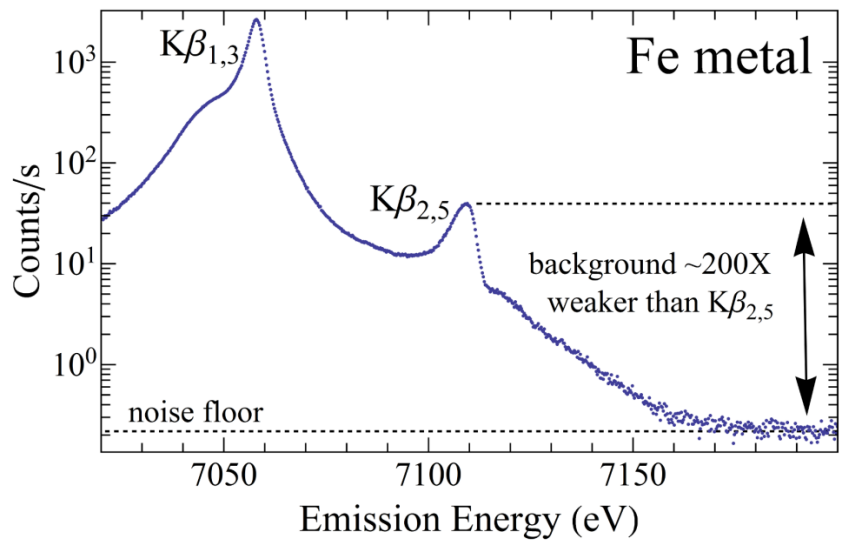

**Figure 1**. A representative Kβ spectrum for Fe metal. Data is extended well above the Fe K-edge (7112 eV) to identify the noise floor. From these measurements we see the background counts (~0.2/s) are 200 x times below than the weak Kβ$_{2,5}$ VTC feature.

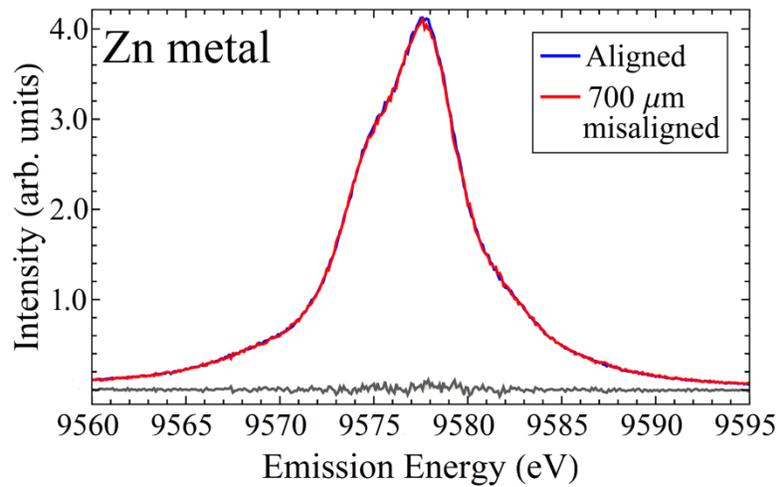

**Figure 2**. Normalized Zn Kβ$_{1,3}$ spectra at the optimal and an extreme misaligned (700 μm) sample location. Also shown is the residual intensity between the two curves (gray line). In this absence of the on-circle, 0.5-mm wide entrance slit, this misalignment would correspond to a relative shift of ~900 meV. Here, the two spectra agree so well as to be nearly indistinguishable.

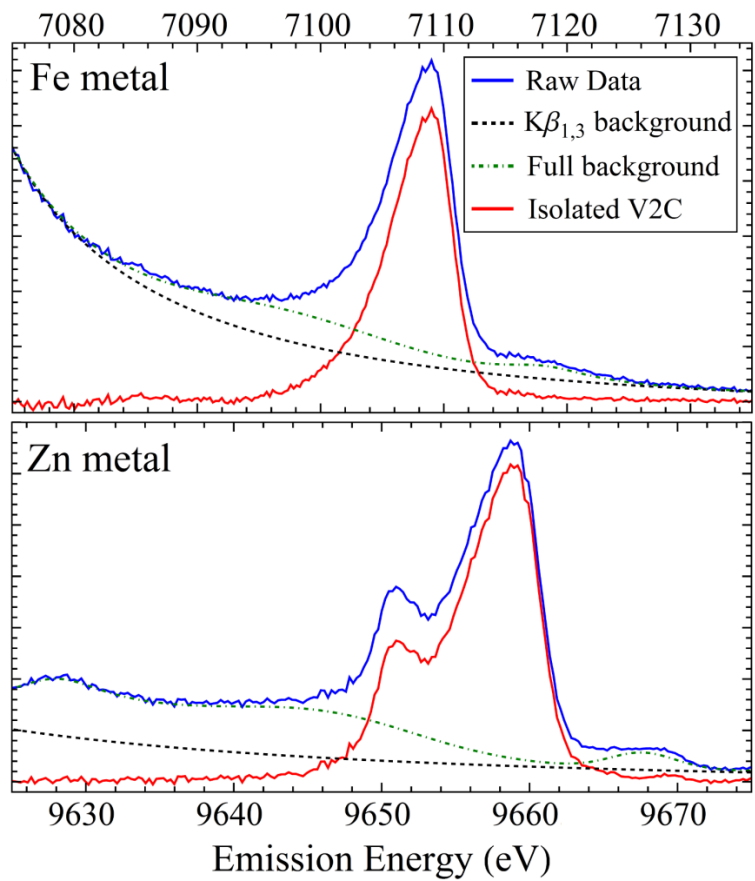

**Figure 3**. The raw (blue) and isolated (red) spectra for Fe and Zn metal VTC spectra. Each full Kβ spectra is fit to a series of pseudo-Voigt functions and the contributions from the K$\beta_{1,3}$ lines (dashed black), elastic scatter, radiative Auger emission, and multi-electron excitations are removed to isolate the VTC features. These background contributions are shown collectively as the dashed-green lines.

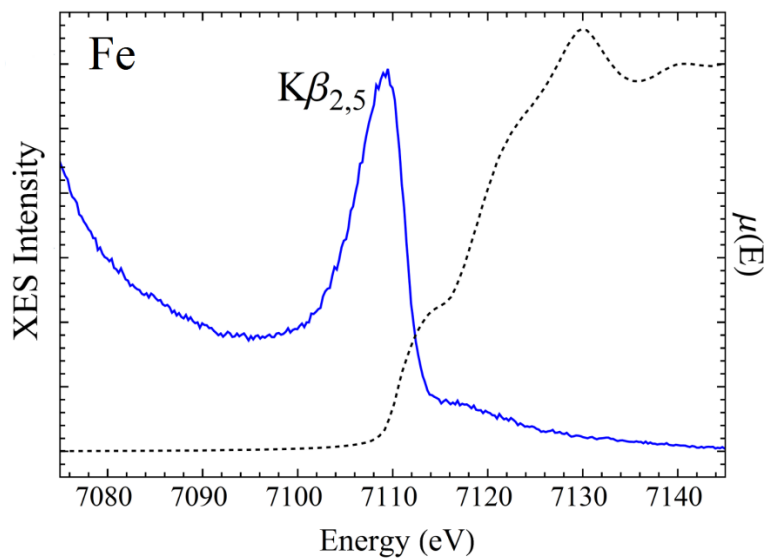

**Figure 4**. The VTC emission (solid, blue) and absorption coefficient (dashed, black), $\mu(E)$, of metallic Fe. Here the K$\beta_{2,5}$ emission feature straddles the rising K-edge absorption resulting in distortions in the spectral shape.

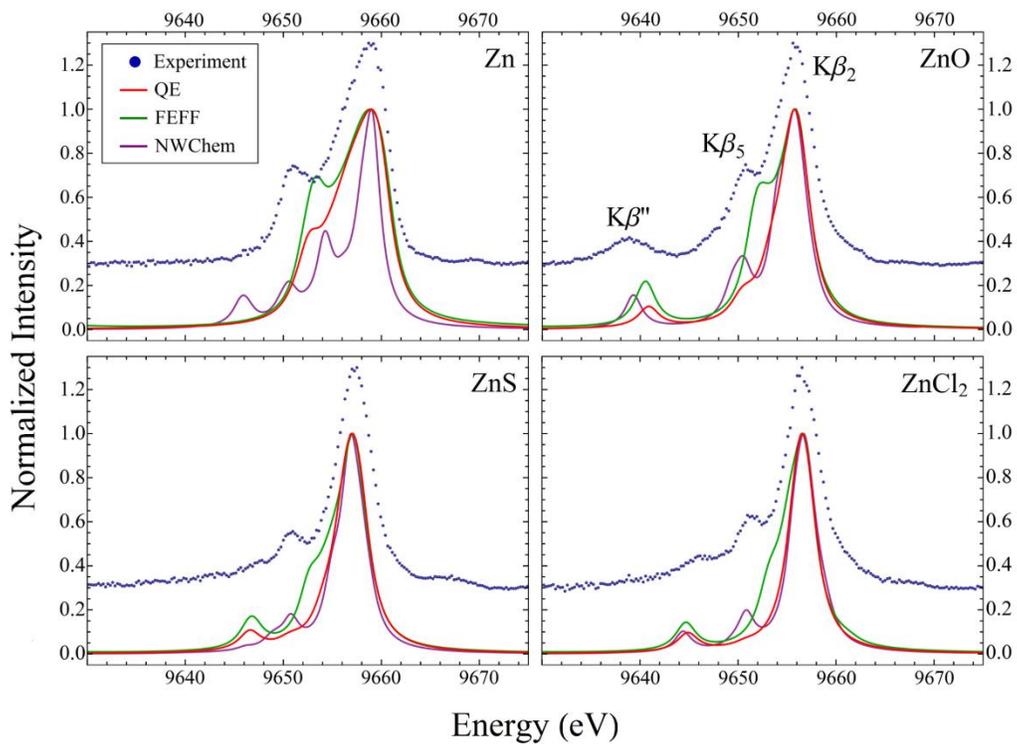

**Figure 5**. Experimental (blue dots) and calculated Quantum ESPRESSO (red), FEFF (green), and NWChem (purple) valence-to-core spectra for various Zn compounds. For comparison, the theoretical results have been broadened as described in the text and shifted to align with the main peak. Experimental data has been offset as indicated.

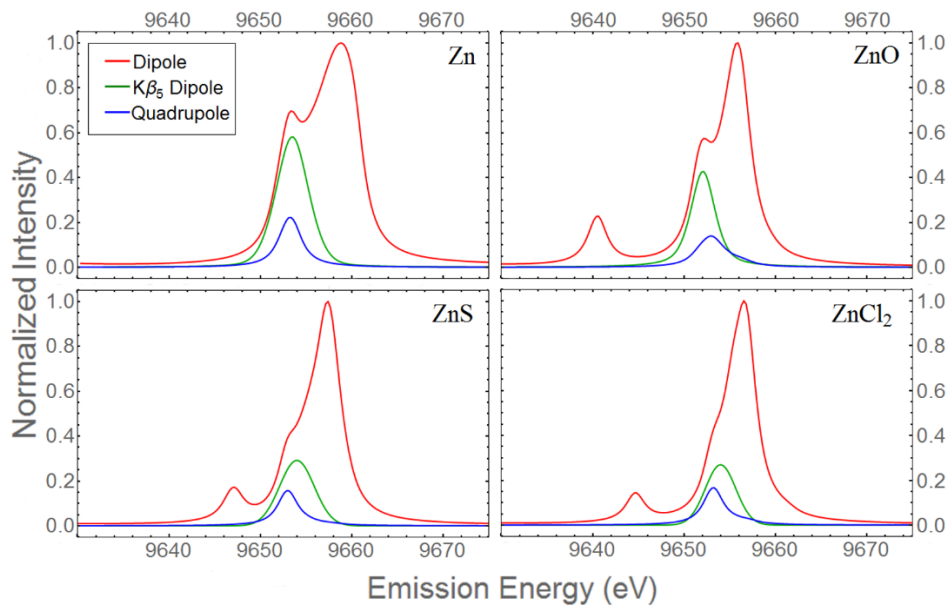

**Figure 6**. The full dipole (red), isolated K$\beta_5$-dipole term (green) and quadrupole (blue) contributions to the FEFF calculation for ZnO. These calculations indicate the K$\beta_5$ term originates from states of both metal 4*p* and 3*d* character.

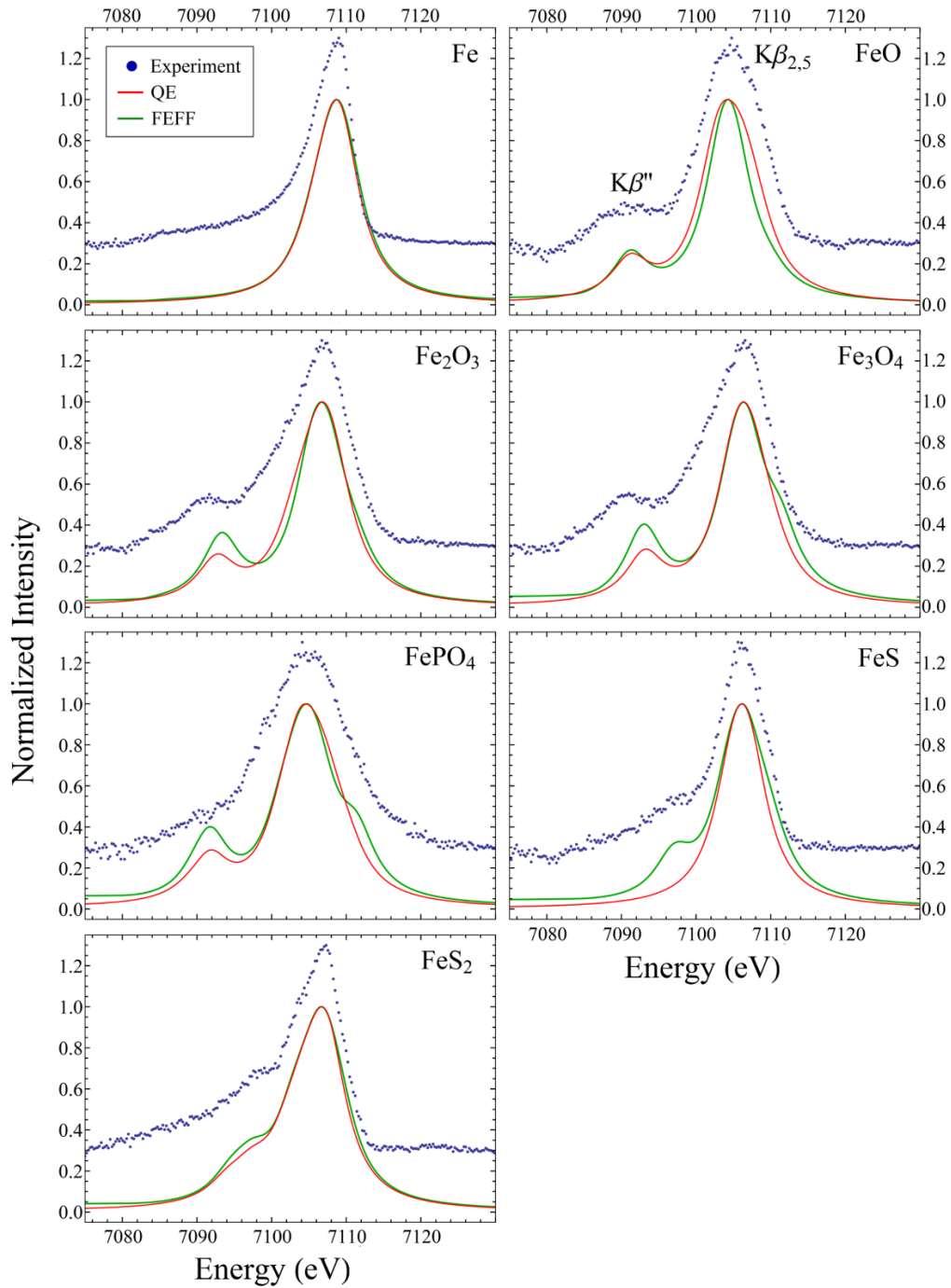

**Figure 7**. Experimental (blue dots) and calculated Quantum ESPRESSO (QE, red) and FEFF (green) valence-to-core spectra for various Fe compounds. For comparison, the theoretical results have been broadened as described in the text and shifted to align with the main peak. Experimental data has been offset as indicated.

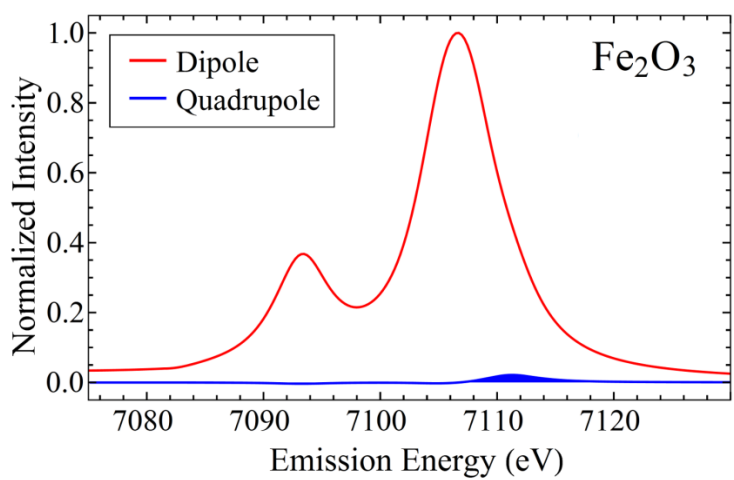

**Figure 8**. The dipole (red) and quadrupole (blue) contributions to the FEFF calculation for $Fe_2O_3$. For emphasis, the quadrupole term is shown with filling to the axis.